\documentclass{IEEEtran4PSCC}
\ifCLASSINFOpdf
   \usepackage[pdftex]{graphicx}
\else
   \usepackage[dvips]{graphicx}
\fi
\usepackage[cmex10]{amsmath}

\usepackage{amsmath,amsfonts}
\usepackage{amsthm}
\usepackage{amssymb}

\usepackage{multicol}
\usepackage{bm}
\usepackage{booktabs}
\usepackage{multirow}

\newtheorem{remark}{Remark}

\newtheorem{problem}{Problem}

\allowdisplaybreaks

\usepackage{float}
\floatstyle{ruled}
\newfloat{problem}{thp}{lop}
\floatname{problem}{Problem}

\usepackage{xcolor}
\definecolor{niceblue}{rgb}{0, 0.5, 1.0}
\definecolor{niceblue}{rgb}{0.125, 0.406, 0.852}
\usepackage{hyperref}
\hypersetup{
    colorlinks=true,
    linkcolor=niceblue,
    filecolor=magenta,      
    urlcolor=niceblue,
    citecolor=niceblue}
    
\usepackage{xcolor}

\makeatletter
\let\old@ps@headings\ps@headings
\let\old@ps@IEEEtitlepagestyle\ps@IEEEtitlepagestyle
\def\psccfooter#1{%
    \def\ps@headings{%
        \old@ps@headings%
        \def\@oddfoot{\strut\hfill#1\hfill\strut}%
        \def\@evenfoot{\strut\hfill#1\hfill\strut}%
    }%
    \def\ps@IEEEtitlepagestyle{%
        \old@ps@IEEEtitlepagestyle%
        \def\@oddfoot{\strut\hfill#1\hfill\strut}%
        \def\@evenfoot{\strut\hfill#1\hfill\strut}%
    }%
    \ps@headings%
}
\makeatother

\psccfooter{%
        \parbox{\textwidth}{\hrulefill \\ \small{24th Power Systems Computation Conference} \hfill \begin{minipage}{0.2\textwidth}\centering \vspace*{4pt} \includegraphics[scale=0.06]{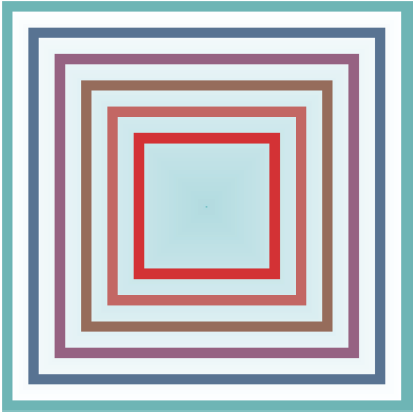}\\\small{PSCC 2026} \end{minipage} \hfill \small{Limassol, Cyprus --- June 8 -- June 12, 2026}}%
}

\newcommand{\btt}[1]{{\fontfamily{lmtt}\selectfont #1}}

\begin{document}
%

\twocolumn

\title{Maximal Load Shedding Verification for Neural Network Models of AC Line Switching}



\author{\IEEEauthorblockN{
Samuel Chevalier\IEEEauthorrefmark{1},
Duncan Starkenburg\IEEEauthorrefmark{1},
Robert Parker\IEEEauthorrefmark{2},
Noah Rhodes\IEEEauthorrefmark{2}
\IEEEauthorblockA{\IEEEauthorrefmark{1} Department of Electrical and Biomedical Engineering\\
University of Vermont,
Burlington, Vermont, USA}
\IEEEauthorblockA{\IEEEauthorrefmark{2} Los Alamos National Laboratory\\
Los Alamos, New Mexico, USA}
}
}

\maketitle

\begin{abstract}

Solving for globally optimal line switching decisions in AC transmission grids can be intractability slow. Machine learning (ML) models, meanwhile, can be trained to predict near-optimal decisions at a fraction of the speed. Verifying the performance and impact of these ML models on network operation, however, is a critically important step prior to their actual deployment. In this paper, we train a Neural Network (NN) to solve the optimal power shutoff line switching problem. To assess the worst-case load shedding induced by this model, we propose a bilevel attacker-defender verification approach that finds the NN line switching decisions that cause the highest quantity of network load shedding. Solving this problem to global optimality is challenging (due to AC power flow and NN nonconvexities), so our approach exploits a convex relaxation of the AC physics, combined with a local NN search, to find a guaranteed lower bound on worst--case load shedding. These under-approximation bounds are solved via \btt{MathOptAI.jl}. We benchmark against a random sampling approach, and we find that our optimization-based approach always finds larger load shedding. Test results are collected on multiple PGLib test cases and on trained NN models which contain more than 10 million model parameters. 

\end{abstract}



\thanksto{\noindent Submitted to the 24th Power Systems Computation Conference (PSCC 2026). LA-UR-25-30627}

\section{Introduction}

As power grid computational problems continue to grow in complexity (e.g., the most recent ARPA-E Grid Optimization test cases, the largest of which contain billions of variables~\cite{holzer2025go}), Machine Learning (ML) is showing increased promise in its ability to provide high-quality solution predictions on very short time scales~\cite{kotary2021end}. Neural Networks (NNs) in particular have been used to predict solutions to challenging Optimal Power Flow (OPF), Unit Commitment (UC), Optimal Transmission Switching (OTS) problems~\cite{chen2024compact,bugaje2023real,acopfRahul}. 

Recently, NNs have also been employed to encode otherwise intractable constraints (e.g., transient stability and bilevel bidding decisions) into various optimization formulations~\cite{garcia2025transient,bilevelElea,murzakhanov2020neural}. Due to their size and nonconvexity, optimizing over NNs is computationally challenging in itself. However, new tools are allowing researchers to directly embed pre-trained ML models into algebraic optimization frameworks. For example, Gurobi and SCIP have recently released Gurobi Machine Learning~\cite{gurobi-machinelearning} and PySCIPOpt-ML~\cite{PySCIPOpt-M}, respectively, for globally optimizing over MILP transformations of pre-trained Neural Networks. Taking a slightly different approach, the JuMP~\cite{DunningHuchetteLubin2017} developers have recently released \btt{MathOptAI.jl}~\cite{dowson2025MathOptAI}, which allows users to optimize over machine learning predictors in a potentially gray-box fashion. For example, using Julia's Python interface, \btt{MathOptAI.jl} uses a Pytorch API to treat a NN like a ``gray-box", calling NN Hessian and gradient information, computed via Pytorch on the GPU, and embedding this information into a JuMP optimization framework. Initial results show that interior point methods can efficiently optimize over NNs with hundreds of millions of parameters using this formulation \cite{parker2025gpu}.

While optimizing over a trained NN can be useful for capturing otherwise intractable constraints, it can also be used for finding adversarial NN inputs and for assessing the performance limits of a given model. Given a set of bounded inputs, the field of ML performance verification seeks to determine if a ML model is capable of reaching some ``unsafe" output state~\cite{brix2023first}. NN verification has been applied to power system operational problems to e.g., verify AC-OPF prediction fidelity~\cite{acopfRahul}, find worst-case DC-OPF optimality gaps~\cite{Venzke:2020}, and determine power flow solution prediction errors~\cite{chevalier2022global}. In this paper, we pose a verification-inspired problem which seeks to determine the maximal amount of load that might be shed if a trained NN is allowed to make Optimal Power Shutoff switching decisions. 

The Optimal Power Shutoff (OPS) problem seeks to find line switching decisions that minimize the risk of wildfire ignitions while minimizing load shed. This problem is introduced in~\cite{rhodes2020balancing}, where authors used a DC linear approximation of the power flow equations. While the DC approximation can lead to highly suboptimal switching decisions, Second-Order-Cone (SOC) convex relaxations of the power equations can find solutions that are very close to the original nonconvex AC power flow solutions~\cite{haag2024long}. A reformulation of the SOC problem introduced in~\cite{rhodes2025second} enables a 4x acceleration the mixed integer SOC switching problem, but it is still not possible to solve the SOC-OPS problem to optimality on a 118-node system in less than 15 hours. This formulation does not yet scale to real-world systems where a grid operator has less than a day to make decisions on where to de-energize power lines. Accordingly, in this paper, we train a NN to predict line switching solutions which solve the SOC-OPS problem. Trained in a supervised fashion, the resulting NN model, which we refer to as an ``OPS-proxy"~\cite{proxyChen} can make high quality switching decision on very short time scales. 

If the OPS-proxy were to be deployed in a control room, a key operational question could be, ``how much load might be shed if the OPS-proxy were allowed to make line switching decisions?" Or, in the most extreme case, ``what is the worst-case load shed that results from actions suggested by the OPS-proxy?" Such load shedding might be the result of ($i$) poor model training, or ($ii$) adverse operating conditions. Regardless, however, the tools of performance verification can be used to solve this problem, which we refer to as the maximal load shedding verification problem. This problem is similar in spirit to the network interdiction problem~\cite{interdiction}, which seeks the worst case load shedding for a minimal set of line losses. However, in our problem, the loads are non-constant, and the switching decisions are controlled by a NN model. Like the network interdiction problem, ours is a hard problem due to its inherent bi-level nature: whenever the OPS-proxy (i.e., the attacker) makes a switching decision, generators can be redispatched by the control room (i.e., the defender) to minimize load-shed. To overcome this complexity, we pose a bilevel verification problem, where the inner nonconvex redispatch problem can be ($i$) convexified and ($ii$) dualized to yield a single-level maximization problem; we formulate this problem with \btt{MathOptAI.jl} and obtain local solutions with IPOPT.

The ML verification literature typically focuses on characterizing the input/output behavior of constrained computational graphs~\cite{wang2021beta}. We propose a verification formulation of a fundamentally different nature. Our formulation asks: what is the worst case damage that an attacker can do, given the defender can optimally defend against the attack? We then apply this formulation in the practical setting of maximizing load shedding. Our specific contributions follow:
\begin{enumerate}
    \item We propose a bilevel attacker-defender formulation of the maximal load shedding verification problem, where an inner ``defender" minimizes load shed through generation redispatch, and an outer ``attacker" maximizes load shed by choosing NN inputs that switch lines.
    \item By treating switching decisions as fixed inputs from the attacker, we dualize the convexified AC redispatch problem to engender a single-level maximization problem.
    \item We formulate the associated verification problem using \btt{MathOptAI.jl} and its embedded PyTorch wrapper, enabling IPOPT to rapidly converge to a guaranteed lower bound on the true verification solution.
\end{enumerate}

The remainder of this paper is structured as follows. In Sec.~\ref{OPS}, we review the Optimal Power Shutoff and AC Feasibility Restoration problems, and we explain how NNs can be used to predict line switching decisions. In Sec.~\ref{Verification}, we introduce the maximal load shedding verification problem, and we propose a relaxed bilevel formulation to bound the solution to this problem. Test results are provided in Sec.~\ref{study}, and conclusions and future work suggestions are presented in Sec.~\ref{conclusion}.

\section{Optimal Power Shutoff Model}\label{OPS}
In this section, we model the Optimal Power Shutoff (OPS) problem. Given a switching decision, we then describe the AC feasibility restoration problem, which redispatches generators to minimize load shed. Finally, we show how NN proxies can be used to predict solutions to the OPS problem.

\subsection{Second Order Cone Optimal Power Shutoff}
The Second Order Cone Optimal Power Shutoff problem (SOC-OPS) from \cite{rhodes2025second} uses a Second Order Cone relaxation of the power flow equations to create a convex relaxation of the original OPS problem.
Consider a power network with components including buses $b \in \mathcal{B}$, power lines $l \in \mathcal{L}$, generators $g \in \mathcal{G}$, shunts $s \in \mathcal{S}$ and nodal power demand $d \in \mathcal{D}$.  
The objective function of the SOC-OPS problem in \eqref{eq:obj} seeks to serve as much of the load demand as possible while minimizing wildfire risk. The multi-objective is
\begin{equation}
    \max \;\; (1-{{\alpha}})\frac{\sum_{d\in\mathcal{D}} x_d {P}^D_d}{{P}^D_{\rm tot}} - {{\alpha}}\frac{\sum_{ij\in\mathcal{L}}z_{ij} {R}_{ij}}{{R}_{\rm tot}} . \label{eq:obj}
\end{equation}

The first term in \eqref{eq:obj} multiplies the power demand of each load ${P}^D_d$  by the continuous status variable $x_d$ of the node, representing the proportion between $0\%$ and $100\%$ of demand that is served.  
The second term in \eqref{eq:obj} considers the wildfire risk. Each energized line contributes to wildfire risk according to the risk of the line ${R}_{ij}$. The risk is multiplied by the binary status variable $z_{ij}$, indicating that a de-energized line $z_{ij}=0$ is no longer at risk of a fault that can ignite a wildfire fire.  
The operator-chosen tradeoff parameter ${\alpha} \in [0,1]$ determines if solutions are more sensitive to wildfire (${\alpha} \xrightarrow{} 1$) or more sensitive to load shed (${\alpha} \xrightarrow{} 0$).  

In the OPS problem, 
$z_{ij}$ is a binary variable indicating the energization state of line $ij$.  Variables $x_d$ and $x_s$ are continuous variables representing continuous load shedding of demand and shunts respectively:
\begin{subequations}
    \begin{align}
        & z_{ij} \in \{0,1\} & \forall ij \in \mathcal{L} \label{eq:zij_bound} \\
        & 0 \le x_d \le 1 & \forall d \in \mathcal{D} \label{eq:xd_bound} \\
        & 0 \le x_s \le 1 & \forall s \in \mathcal{S} . \label{eq:xs_bound} 
    \end{align}\label{eq:z_bounds}
\end{subequations}

Generator limits are enforced by eq. \eqref{eq:ops_active_gen_limits} and \eqref{eq:ops_reactive_gen_limits}.  Real power $P^G_g$ from generator $g$ is constrained to be between $0$ and the maximum generator limit; the lower limit is relaxed to $0$: 
\begin{align}
        0 &\le P_{g}^G \le  \overline{{P}^G_g} \quad \forall g \in \mathcal{G}, \label{eq:ops_active_gen_limits}\\
        \underline{{Q}^G_g} &\le Q_{g}^G \le  \overline{{Q}^G_g} \quad \forall g \in \mathcal{G} .\label{eq:ops_reactive_gen_limits}
\end{align}

Nodal real and reactive power balance are constrained in eqs. \eqref{eq:ops_soc_active_power_balance} and \eqref{eq:ops_soc_reactive_power_balance}. At each node $i$, the sum of generator power $P^G_g$, net flow on lines $P^L_{ij}$, served load $x_d {P}^D_d$, and shunt power ${g}_i W^S_i$ must sum to zero, where ${g}_i$ is the shunt conductance at node $i$ and $W^S_i$ is the squared voltage magnitude variable seen by the shunt.  Reactive power is similarly constrained for reactive generator power $Q^G_g$, reactive line power $Q^L_{ij}$, reactive power demand $x_d Q^D_d$, and reactive shunt power ${b}_s W^S_s$ where ${b}_s$ is the node $i$ shunt susceptance and $W^S_s$ is the shunt voltage:
\begin{equation}
        \sum_{g\in\mathcal{B}_i^\mathcal{G}}P_{g}^G - \!\!\!\!\!\! \sum_{(i, j)\in\mathcal{B}_i^\mathcal{L}} \!\!\!\!\!  P_{i j}^L - \!\!\!\!  \sum_{d\in\mathcal{B}_i^\mathcal{D}} \!\!\! x_{d} {P}^D_d  - \!\!\!\!\sum_{s\in\mathcal{B}_i^{\mathcal{S}}}{g}_s W^S_i = 0 \quad \forall i \in \mathcal{B},  \label{eq:ops_soc_active_power_balance} 
\end{equation}
\begin{equation}
        \sum_{g\in\mathcal{B}_i^\mathcal{G}}Q_{g}^G - \!\!\!\!\!\! \sum_{(i, j)\in\mathcal{B}_i^\mathcal{L}} \!\!\!\!\!  Q_{i j}^L - \!\!\!\!  \sum_{d\in\mathcal{B}_i^\mathcal{D}} \!\!\! x_{d} {Q}^D_d + \!\!\! \sum_{s\in\mathcal{B}_i^{\mathcal{S}}}{b}_s W^S_i = 0 \quad \forall i \in \mathcal{B}. \label{eq:ops_soc_reactive_power_balance} 
\end{equation}

The thermal limit of a power line, given as ${{T}_{ij}}$ in eq. \eqref{eq:ops_complex_thermal_limit}, is the maximum apparent power that can flow across the line.  However, if the line is de-energized, the maximum flow across the line is $0$, as controlled by line energization state $z_{ij}$:
\begin{subequations}
\begin{align}
        (P_{i j}^L)^2 + (Q_{i j}^L)^2 &\le {T}_{i j}^2z_{i j} \quad \forall ij \in \mathcal{L} \label{eq:ops_complex_thermal_limit_to} \\
        (P_{j i}^L)^2 + (Q_{j i}^L)^2 &\le {T}_{i j}^2z_{i j} \quad \forall ij \in \mathcal{L}. \label{eq:ops_complex_thermal_limit_fr}
\end{align} \label{eq:ops_complex_thermal_limit}
\end{subequations}

The Second-Order-Cone (SOC) relaxation of the power flow equations replace squared voltage variables with lifted variables, like $W_{ii}$, to replace $V_iV_i$ from the AC power flow equations.  Eq.~\eqref{eq:ops_soc_voltage_bus} provides the bounds on $W_{ii}$:
\begin{equation}
        {\underline{V}}^2_i \le  W_{ii}  \le  {\overline{V}}^2_i \quad  \forall i \in \mathcal{B} . \label{eq:ops_soc_voltage_bus}
\end{equation}

Additional variables are added to represent the voltage at each end of the power line. The variables $W^{To}_{ij}$ and $W^{Fr}_{ij}$ are equivalent to $W_{ii}$ and $W_{jj}$ when the line is energized, but are constrained to $0$ when the line is de-energized.  These additional variables allow for an improved root relaxation and decrease solve time for topology optimization decisions \cite{rhodes2025second}:  
\begin{subequations}
    \begin{align}
        & z_{ij} {\underline{V}}^2_i \le  W^{Fr}_{ij}  \le  z_{ij} {\overline{V}}^2_i & \forall ij \in \mathcal{L} \\
        & z_{ij} {\underline{V}}^2_j \le  W^{To}_{ij}  \le  z_{ij} {\overline{V}}^2_j & \forall ij \in \mathcal{L}. 
    \end{align} \label{eq:ops_soc_voltage_fr_to}
\end{subequations}

Eq.~\eqref{eq:ops_soc_voltage_fr_to_ii_jj} links the value of $W^{Fr}_{ij}$ and $W_{jj}$ (and $W^{To}_{ij}$ to $W_{ii}$) when the line is energized, but decouple the variables when the line is de-energized:
\begin{subequations}
    \begin{align}
        & W_{ii} \ge W^{Fr}_{ij}  \ge W_{ii} - {\overline{V}}_i^2 (1-z_{ij}) & \forall ij \in \mathcal{L} \\
        & W_{jj} \ge W^{To}_{ij}  \ge W_{jj} - {\overline{V}}_j^2 (1-z_{ij}) & \forall ij \in \mathcal{L}. 
    \end{align}  \label{eq:ops_soc_voltage_fr_to_ii_jj}
\end{subequations}
    Variables $W^R_{ij}$ and $W^I_{ij}$ represent the nonlinear real and imaginary projections of the voltage terms $V_iV_j \cos{(\theta_i-\theta_j)}$ and $V_iV_j \sin{(\theta_i-\theta_j)}$ from the AC power flow equations.  The bounds for these variables can be computed from the voltage magnitude limits and voltage angle limits, with details found in \cite{coffrin2015qc}. Like the other voltage variables representing a line, these variables are constrained to $0$ when the line is de-energized:
\begin{subequations}
    \begin{align}
& z_{ij}{\underline{W}}^R_{ij} \le  W^R_{ij}  \le  z_{ij}{\overline{W}}^R_{ij} \quad \forall ij \in \mathcal{L}  \\
& z_{ij}{\overline{W}}^I_{ij} \le  W^I_{ij}  \le  z_{ij}{\underline{W}}^I_{ij} \quad   \forall ij \in \mathcal{L}.
        \end{align} \label{eq:ops_soc_ij_ri_bounds}
\end{subequations}
The voltage angle limit across a power line is captured by
\begin{equation}
        \tan \left({\underline{\theta}}_{ij}\right) W^R_{ij} \le W^I_{ij} \le \tan\left({\overline{\theta}}_{ij} \right) W^R_{ij} \quad \forall ij \in \mathcal{L}.
    \label{eq:ops_soc_angle_limits}
\end{equation}
The rotated SOC constraint links nodal and line voltages via
 \begin{equation}
     \left(W^R_{ij}\right)^2 +  \left(W^I_{ij}\right)^2  \le W^{Fr}_{ij} W^{To}_{ij}  \quad \forall ij \in \mathcal{L}.\label{eq:ops_soc_cone_perp}
 \end{equation}

Utilizing the voltage variables introduced in \eqref{eq:ops_soc_voltage_bus}-\eqref{eq:ops_soc_cone_perp}, the power flow equations in the SOC relaxation are shown in eq. \eqref{eq:ops_soc_powerflow}, located in the appendix. Finally, eq.~\eqref{eq:ops_soc_shunt_relaxation} introduces the McCormick relaxation of the shunt voltage variable.  The shunt voltage is determined by the voltage variable $W_{ii}$, but we allow continuous shedding of the load at the shunt, $W^S_i = x_s W_{ii}$.  
\begin{subequations}
    \begin{align}
        &  W^S_{s} \ge 0  \quad \forall s \in \mathcal{B}^\mathcal{S}_i,      & \forall i \in \mathcal{B} \\
        &  W^S_{s} \ge {\overline{V}}^2_i (x_s -1 ) + W_{ii}                  & \forall s \in \mathcal{B}^\mathcal{S}_i, \; \forall i \in \mathcal{B} \\
        &  W^S_{s} \le W_{ii}  \quad \forall s \in \mathcal{B}^\mathcal{S}_i, & \forall i \in \mathcal{B} \\
        &  W^S_{s} \le {\overline{V}}^2_i x_s  \quad \forall s \in \mathcal{B}^\mathcal{S}_i, & \forall i \in \mathcal{B}.
    \end{align} \label{eq:ops_soc_shunt_relaxation}
\end{subequations}

The complete optimization problem is shown in SOC-OPS.  
\begin{problem}[h]
\caption{\hspace{-0.1cm}\textbf{:} SOC Optimal Power Shutoff (SOC-OPS)}
\label{model:epi}
\vspace{-0.25cm}
\begin{align}
    \max\quad &  \mbox{Objective \eqref{eq:obj}} \nonumber\\
{\rm s.t.}\quad& \mbox{Energization states:}~\eqref{eq:z_bounds} \nonumber \\[-2pt]
& \mbox{Generation constraints:}~\eqref{eq:ops_active_gen_limits}, \eqref{eq:ops_reactive_gen_limits}  \nonumber\\[-2pt]
& \mbox{SOC power flow:}~\eqref{eq:ops_soc_active_power_balance}\rm{-}\eqref{eq:ops_soc_shunt_relaxation},\eqref{eq:ops_soc_powerflow}.
\nonumber
\end{align}
\vspace{-0.5cm}
\end{problem}

\subsection{Feasibility Restoration via Redispatch}
To ensure the SOC-OPS solution is AC feasible, we solve an AC power flow optimization problem  where the objective function minimizes unserved load with the new topology that comes from the SOC-OPS solution.  No other variables are constrained by the SOC-OPS solution. This results in a continuous, non-convex optimization problem that we solve to local optimality. This problem is called the AC-Redispatch problem~\cite{rhodes2021powermodelsrestoration}. For ease of notation, we reuse some equations from the SOC-OPS problem, but the value of de-energization decisions are fixed to the values ${\hat z}$ from an OPS solution, 
\begin{equation}
    z_{ij} = {\hat{z}}_{ij} \quad \forall ij \in \mathcal{L}. 
    \label{eq:redis_fixed}
\end{equation}
The objective is to minimize load shed (i.e., maximize load delivery), and is equivalent to the first term in \eqref{eq:obj}:
\begin{equation}
    \min \; \sum_{d\in\mathcal{D}} (1-x_d) {P}^D_d  \label{eq:load_objective}
\end{equation}
Nodal voltage variables are used in this formulation and are constrained to be between their voltage limits,
\begin{equation}
    {\underline{V}}_i \le V_i \le {\overline{V}}_i \quad \forall i \in \mathcal{B} \label{eq:ops_voltage_magnitude}.
\end{equation}

The energy balance equations now use the variable $V_i$ rather than the lifted variable $W_{ii}$ for shunt power:
\begin{equation}
        \sum_{g\in\mathcal{B}_i^\mathcal{G}}P_{g}^G - \!\!\!\!\!\! \sum_{(i, j)\in\mathcal{B}_i^\mathcal{L}} \!\!\!\!\!  P_{i j}^L - \!\!\!  \sum_{d\in\mathcal{B}_i^\mathcal{D}} \!\!\! x_{d} {P}^D_d -
        {g}_i V_i^2 x_s = 0 \; \forall i \in \mathcal{B}  \label{eq:ops_active_power_balance} 
\end{equation}
\begin{equation}
        \sum_{g\in\mathcal{B}_i^\mathcal{G}}Q_{g}^G - \!\!\!\!\!\! \sum_{(i, j)\in\mathcal{B}_i^\mathcal{L}} \!\!\!\!\!  Q_{i j}^L - \!\!\!  \sum_{d\in\mathcal{B}_i^\mathcal{D}} \!\!\! x_{d} {Q}^D_d + {b}_i V_i^2 x_s = 0 \; \forall i \in \mathcal{B}  \label{eq:ops_reactive_power_balance} 
\end{equation}

The AC power flow equations are given by \eqref{eq:ops_ac_powerflow} in the appendix. Finally, we introduce a constraint on the voltage angle difference.  The big-M formulation de-couples node pairs where a connecting powerline has been de-energized.
\begin{subequations}
\begin{align}
    & \theta_{i} - \theta_{j} \le \overline{{\theta}_{ij}}+ {\theta}^{\Delta}_{\rm max} (1-z_{i j}) \label{eq:ops_votlage_angle_max} \\
    & \theta_{i} - \theta_{j} \ge -\overline{{\theta}_{ij}}- {\theta}^{\Delta}_{\rm max} (1-z_{i j}). \label{eq:ops_votlage_angle_min}
\end{align} \label{eq:ops_voltage_angle}
\end{subequations}

The final optimization problem is given by AC-Redispatch.
\begin{problem}[h]
\caption{\hspace{-0.1cm}\textbf{:} AC-Redispatch (Feasibility Restoration)}
\label{model:ac_redispatch}
\vspace{-0.25cm}
\begin{align}
    \min \quad &  \mbox{Objective \eqref{eq:load_objective}} \nonumber\\
\mbox{s.t.}\quad
& \mbox{Component relationships:}~\eqref{eq:redis_fixed},\eqref{eq:xd_bound},\eqref{eq:xs_bound} \nonumber \\[-2pt]
& \mbox{Generation constraints:}~\eqref{eq:ops_active_gen_limits},\eqref{eq:ops_reactive_gen_limits}  \nonumber\\[-2pt]
& \mbox{AC power flow:}~\eqref{eq:ops_complex_thermal_limit}
,\eqref{eq:ops_voltage_magnitude} \rm{-} \eqref{eq:ops_voltage_angle}, \eqref{eq:ops_ac_powerflow} \nonumber
\end{align}
\vspace{-0.5cm}
\end{problem}

The solution to the AC-Redispatch problem represents the actual load shed if the topology from the OPS solution were to be implemented, and a gap in load shed between the SOC-OPS and AC-Redispatch optimization problems indicates that the SOC relaxation was not tight.  

\subsection{Neural Network OPS Predictions}
Solving SOC-OPS to global optimality is a hard computational problem, taking more than 15 hours on the 118-bus network. To overcome this computational bottleneck, we train a NN, via supervised learning, to predict OPS switching solutions, i.e., we train the model to predict which transmission lines should be switched on or off to globally maximize \eqref{eq:obj}. Since the NN only makes switching decisions, the decisions are then passed to the AC-Redispatch function, which restores network feasibility while minimizing load shed. The NN inputs are based on network load variations, line risks, and the weighting risk parameter $\alpha$. The output is a vector of line status activation probabilities ${ L} \in {\mathbb R}^{|{\mathcal L}|}$, where ${ L}_{ij} \in [0,1]$ is the output of a logistic function $(1+e^{-x})^{-1}$ and represents the probability that line $ij$ should be activated\footnote{In practical deployment, lines with ${ L}_{ij}\ge0.5$ are predicted by the model to be active, while lines with ${ L}_{ij}<0.5$ are predicted to be deactivated.}:
\begin{align}
{\rm NN}_{{\rm in}} \,& =\gamma \triangleq \{{P}^{D},{Q}^{D},{R},\alpha\}\\
{\rm NN}_{{\rm out}} & =\{{L}\}.
\end{align}

\begin{figure}
\begin{center}
\includegraphics[width=0.98\columnwidth]{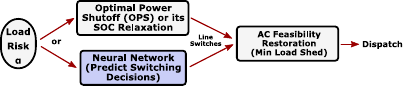}
\caption{Given a set of inputs (grid loads, line risks, and tradeoff parameter $\alpha$), OPS (top path) will make an optimal set of switching decisions, and then those decisions are used in the AC feasibility restoration step. Alternatively, we train a NN to predict optimal switching decisions (bottom path), which are also subjected to feasibility restoration.}\label{fig:ops}
\end{center}
\end{figure}
Thus, the NN maps from $\gamma$ to $ L$ via
\begin{align}
{\rm NN}(\gamma) \rightarrow { L}.
\end{align}
The interaction between NN inputs and outputs are depicted in Fig.~\ref{fig:ops}. The feed-forward NN architectures trained and tested in this paper are sigmoid-based with an input layer, two hidden layers, and an output layer. For each power grid, we train four models of varying sizes (with 32, 128, 512, and 2048 activation functions per hidden layer). The total number of trainable NN parameters, which is a function of power grid size (since input/output dimensions vary) are in Table \ref{tab:nn_sizes}.

\begin{table}
   \caption{Number of NN Parameters Across 12 Trained NNs} 
   \label{tab:nn_sizes}
   \small
   \centering
   \begin{tabular}{c|cccc}
   \toprule\toprule
   \multirow{2}{4em}{\centering\textbf{PGLib\\Case}} & \multicolumn{4}{c}{Hidden Layer Size} \\ &32  & 128  & 512  & 2048 \\  

   \midrule 
   $\;$\textit{14\_ieee} & 4.37$k$ & 42.00$k$ & 561.17$k$  & 8.54$M$ \\
   $\;$\textit{24\_ieee} & 6.18$k$ & 49.19$k$  & 589.86$k$  & 8.65$M$ \\
   $\;$\textit{118\_ieee}& 21.82$k$ & 111.29$k$ & 837.81$k$  & 9.64$M$\\
\bottomrule
   \end{tabular}
\end{table}

\section{Model Verification}\label{Verification}
As discussed in the previous subsection, a trained NN will feed a line switching decision into an AC redispatch function, which minimizes load shed (NN $\rightarrow$ redispatch $\rightarrow$ total load shed). Given this framework, in this section, we introduce the central verification problem which this paper considers: maximal load shedding. In particular, we search for NN switching decisions which will lead to the highest amount of load being shed from the power system. This is a nontrivial problem: simply ``turning up the load" may indeed lead to higher levels of load shed; however, as load increases, the NN will be incentivized to keep the lines which serve the growing load fully energized, in service of implicitly maximizing objective \eqref{eq:obj}. In this section, we offer an optimization-based framework for finding adverse network configurations (i.e, NN inputs) which lead to situations of maximal load shedding.

\subsection{Attacker-Defender Model}

To model the maximal load shedding verification problem, we pose an attacker-defender style bilevel optimization problem, where the outer ``attacker" attempts to choose network loading configurations which lead to maximized load shedding through NN switching decisions. Meanwhile, the inner ``defender" redispatches generation to minimize the load shedding:
\begin{align}\label{eq: attack_defend}
\text{\textbf{bilevel} \textbf{verification}: }\;\;\,\overbrace{\max_{p_d} \;\; \underbrace{\min_{p_g}\;\;\{\text{load shed}\}}_{\text{\textbf{defender} (redispatch)}}}^{\text{\textbf{attacker} (choose network loading)}}.
\end{align}
In \eqref{eq: attack_defend}, the ``$p_d$" decision variables represent NN inputs (generally, network loading, but also line risks), and ``$p_g$" represents control actions (generally, generator redispatch).

In addition to the bilevel nonconvexity, there are two other sources on nonconvexity in \eqref{eq: attack_defend}: ($i$) the AC power flow equations in the redispatch problem, and ($ii$) the neural network mapping. As a true bilevel optimization problem, \eqref{eq: attack_defend} is very challenging to solve, even locally. In order to state this problem explicitly, we first \textit{canonicalize} the primal redispatch problem: 
\begin{subequations}\label{eq: canon}
\begin{align}
\min_{x}\quad & h^{T}x\\
{\rm s.t.}\quad & Ax+b\left(\gamma,{L}\right)=0\\
 & Cx+d\left(\gamma,{L}\right)\le0\\
 & f(x)=0,
\end{align}
\end{subequations}
where $x$ is the vector of primal variables in AC-Redispatch, while $\gamma$ and ${L}$ represent the network loading and line status information that are exogenous to, but necessary for solving, the load shedding minimization problem. We do not provide the full canonicalization in this paper, but it can be directly constructed from the previously presented constraints.

\begin{remark}
    \eqref{eq: canon} is a canonicalized form of the AC-Redispatch (Problem \ref{model:ac_redispatch}). They are otherwise equivalent.
\end{remark}

Assuming a NN has been trained to map between network loading and line status, we embed this NN mapping as an outer level constraint in the maximization problem:
\begin{subequations}\label{eq: bilevel}
\begin{align}
s^{\star}=\max_{\substack{\underline{\gamma}\le\gamma\le\overline{\gamma}\\ {\rm NN}\left(\gamma\right) = { L}}} \;\min_{x}\quad & h^{T}x\\
{\rm s.t.}\quad & Ax+b\left(\gamma,{L}\right)=0\\
 & Cx+d\left(\gamma,{L}\right)\le0\\
 & f(x)=0.\label{eq: nonconvex}
\end{align}
\end{subequations}
The global solution to \eqref{eq: bilevel}, which has three sources of nonconvexity, will yield the largest load shedding that can be caused by NN switching predictions. Notably, we bound the NN inputs $\gamma$ by the upper and lower bounds $\underline{\gamma}$, $\overline{\gamma}$. This represents the range over which we wish to verify the performance of the NN. In this paper, we test over ranges given by
\begin{align}\label{eq: gamma_bounds}
\underline{\gamma}=\left[\begin{array}{c}
0.75{P}^{D,0}\\
0.75{Q}^{D,0}\\
0.25\\
0.25
\end{array}\right],\;\overline{\gamma}=\left[\begin{array}{c}
1.25{P}^{D,0}\\
1.25{Q}^{D,0}\\
0.75\\
0.75
\end{array}\right].
\end{align}

\subsection{Inner OPF Relaxation and Dualization}
The only nonconvexity in the inner minimization problem of \eqref{eq: bilevel} comes from AC power flow physics. To make \eqref{eq: bilevel} more tractable, we perform a convex relaxation for the nonlinearity in $f(x)$, yielding $x\in {\mathcal K}$, where ${\mathcal K}$ is the intersection of rotated second order and linear cones\footnote{The conic constraints are technically parameterized by network parameters and line status variables, as shown, for example, in constraint \eqref{eq:ops_complex_thermal_limit}. With slight abuse of notation, however, we represent all conic constraints via $x\in {\mathcal K}$. }:
\begin{subequations}\label{eq: bi_cvx}
\begin{align}
{\tilde s} = \max_{\substack{\underline{\gamma}\le\gamma\le\overline{\gamma}\\ {\rm NN}\left(\gamma\right) = { L}}}\;\min_{x}\quad & h^{T}x\\
{\rm s.t.}\quad & Ax+b\left(\gamma,{L}\right)=0\\
 & Cx+d\left(\gamma,{L}\right)\le0\\
 & x\in\mathcal{K}.
\end{align}
\end{subequations}
This convexification will result into a lower bound on the true load shedding solution. In this paper, we exclusively study the SOC relaxation of the power flow equations. However, other convex relaxations, with any combination of tightening cuts, are also admissible in our framework.

\begin{remark}\label{eq: remark_bound}
    Due to the convexification of \eqref{eq: nonconvex}, ${\tilde s} \le s^{\star}$. Stated alternatively, \eqref{eq: bi_cvx} will always lower bound the worse case load shedding.

    \begin{proof}
        For any feasible value of $\gamma$ and $ L$, denote $x^{\star}$ and ${\tilde x}$ as the solutions which minimize the inner primals \eqref{eq: bilevel} and \eqref{eq: bi_cvx}, respectively. Since ${\tilde x}$ is the solution of a relaxation, the corresponding minimization will lower bound the minimization provided by $x^{\star}$. Since this is true for any values of $\gamma$ and $ L$, it must also be true for their optimal values. Therefore, ${\tilde s} \le s^{\star}$ must hold.
    \end{proof}
\end{remark}

To solve \eqref{eq: bi_cvx}, we take the dual of the inner minimization, which is convex in the primal variable $x$. To take the dual, we first build the Lagrangian associated with the system:
\[
\mathcal{L}=h^{T}x+\lambda^{T}\left(Ax\!+\!b\left(\gamma,{L}\right)\right)+\mu^{T}\left(Cx\!+\!d\left(\gamma,{L}\right)\right)-s^{T}x
\]
where $s$ is a set of dual conic variables constrained to the intersection of dual cones: $s\in {\mathcal K}^*$. The unconstrained primal will be bounded only when $h^{T}+\lambda^{T}A+\mu^{T}C-s^{T}=0$. Capturing this equality constraint, the associated dual program is given by
\begin{subequations}
\begin{align}
\max_{\substack{\underline{\gamma}\le\gamma\le\overline{\gamma},\\
{\rm NN}\left(\gamma\right)={{L}}
}
}\;\max_{\mu\ge0,\lambda,s\in\mathcal{K}^{*}}\quad & \lambda^{T}b\left(\gamma,{L}\right)+\mu^{T}d\left(\gamma,{L}\right)\\
{\rm s.t.}\quad & h^{T}+\lambda^{T}A+\mu^{T}C-s^{T}=0.
\end{align}
\end{subequations}
Finally, due to the associativity of the maximum operator, we can write this as a single maximization problem:

\begin{problem}[h]
\caption{\hspace{-0.1cm}\textbf{:} Lower Bound on Maximal Load Shedding}
\label{model:max_shed}
\vspace{-0.25cm}
\begin{subequations}\label{eq: max_daul}
\begin{align}\max_{\mu\ge 0 \lambda,\gamma,s}\quad & \lambda^{T}b\left(\gamma,L\right)+\mu^{T}d\left(\gamma,L\right)\\
{\rm s.t.}\quad\;\; & h^{T}+\lambda^{T}A+\mu^{T}C-s^{T}=0\\
 & {\rm NN}\left(\gamma\right)=L\\
 & \underline{\gamma}\le\gamma\le\overline{\gamma}\\
 & s\in\mathcal{K}^{*}
\end{align}
\end{subequations}
\vspace{-0.5cm}
\end{problem}

We use IPOPT to find a local solution to \eqref{eq: max_daul}.

\begin{remark}
    Any feasible, local solution to the maximization problem \eqref{eq: max_daul} will lower bound the original bilevel optimization problem \eqref{eq: bilevel}.
    \begin{proof}
        Since the transformation from \eqref{eq: bi_cvx} to \eqref{eq: max_daul} is exact (under mild technical assumptions which generally hold for power system optimization problems), we may use the proof of Remark \eqref{eq: remark_bound} to guarantee that \eqref{eq: max_daul} lower bounds \eqref{eq: bilevel}.
    \end{proof}
\end{remark}

There are two primary reasons that a local solution to \eqref{eq: max_daul} will only yield a lower bound on the true problem solution. First, the non-convex AC equations were relaxed in the construction of this model, so effects of load shedding will generally be under-represented. Second, the NN provides a nonconvex mapping (which we did not convexify), so there might always exist a set of inputs loading which cause an even worse load shed. Therefore, \eqref{eq: max_daul} must be interpreted as a tractable optimization problem which provides a load shed bound which is ``at least as bad" as the very worst case load shedding.  In other words, it cannot provide a tight guarantee on worst-case load shedding, but it can provide a lower bound.

\subsection{Snapping Switching Decisions}
There is one other source of approximation error embedded in \eqref{eq: max_daul}. Neural networks with continuous activation functions make continuous decisions. Here, the logistic function output from our model returns the normalized probability ${ L}_{ij}$, between 0 and 1, that a line will be active or not. This probability can be interpreted as the binary relaxation of the line status variable $z_{ij} \in [0,1]$.

\begin{figure}
\begin{center}
\includegraphics[width=0.98\columnwidth]{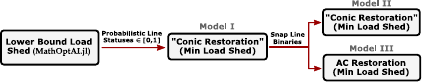}
\caption{Initially, we lower-bound the load shed by solving \eqref{eq: max_daul}. Model I, solved via Gurobi, uses the relaxed line statuses found by \eqref{eq: max_daul} to solve the convex AC redispatch problem. After snapping the binaries, Model II, solved via Gurobi, solves a convex relaxation of the AC feasibility restoration problem, and Model III, solved via IPOPT, solves the true AC feasibility restoration problem.}\label{fig:snapping}
\end{center}
\end{figure}

In practical application, we could append a step function $u_s(\cdot)$ on the output of the NN (e.g, $u_{s}({L}_{ij}-0.5)\in\{0,1\}$) to snap switching decision to their binary values. However, verifying (i.e., optimizing) over $u_s(\cdot)$ is generally challenging, since the function provides no useful sensitivity information. 
To overcome this challenge, we verify the original NN, whose final layer consists of sigmoid activation functions. After solving \eqref{eq: max_daul} (see Fig.~\ref{fig:snapping}), we then fix the line statuses and re-solve the network using a conic solver (Gurobi) to ``clean up" the solution (see Appendix \ref{AppB}). We call this Model I, which we refer to as a ``conic restoration" step, since it optimally solves a convex relaxation of the AC-Redispatch problem. Next, we snap the line statuses according to the step function $z_{ij} = u_{s}({L}_{ij}-0.5)\in\{0,1\}$). Via IPOPT, we then solve the true AC restoration problem with binary line statuses in Model III. However, since this is a nonconvex problem, we also solve its convex relaxation in Model II, to understand how far the AC redispatch solution is from the global bound.

\section{Case Study}\label{study}
In this section, we present test results collected on three small PGLib~\cite{babaeinejadsarookolaee2019power} test cases. We focus our results on smaller-scale power grid test cases (i.e., where we could collect enough OPS data to train a NN model) and larger-scale NN architectures (the largest test network had almost 10 million model parameters, as shown in Table \ref{tab:nn_sizes}). All simulation code is publicly posted on GitHub\footnote{\url{https://github.com/SamChevalier/LoadShedVerification}}. We used Gurobi~\cite{gurobi} and IPOPT~\cite{wachter2006implementation} for all numerical optimization with JuMP~\cite{Lubin2023} as the algebraic modeling software. All NN models were trained using PyTorch~\cite{paszke2019pytorch}, and \btt{MathOptAI.jl} used the PyTorch API to differentiate, via GPUs, through the NN models.

\subsection{Verification Results}
For all 12 cases (3 PGLib grids x 4 trained NNs each), we solved the sequence of models shown in Fig.~\ref{fig:snapping}. No benchmark solver can solve the full bilevel optimization problem given in \eqref{eq: bilevel}. Instead, we used a sampling based approach to benchmark our results. For all 12 cases, we uniformly sample a $\gamma$ value (i.e., we uniformly sample from the bounds given by \eqref{eq: gamma_bounds}), and we pass this sample into the NN. The NN then returns a vector of probabilistic line statuses, which we applied to the pipeline of models in Fig.~\ref{fig:snapping}. This procedure was repeated 100 times for each model; samples which resulted in numerical infeasibility were rejected and re-sampled; more technical implementation details are provided in Appendix \ref{AppB}.

\begin{figure}
\begin{center}
\includegraphics[width=1\columnwidth]{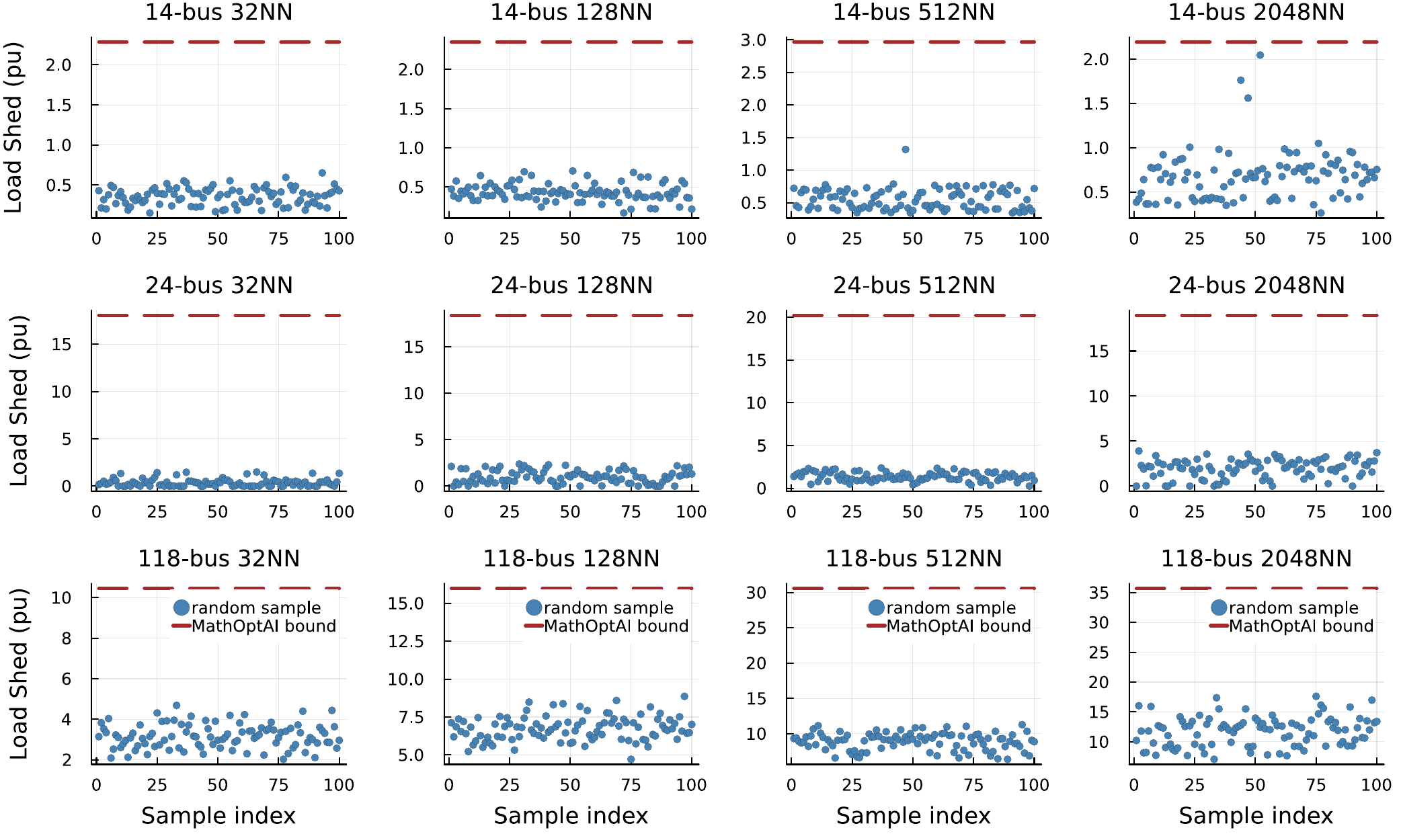}
\caption{Results for Model I in Fig.~\ref{fig:snapping}. For all 12 cases, we plot the load shedding induced by random sampling (blue dots) vs the bound found by \btt{MathOptAI.jl} (red dashed line). In each case, the \btt{MathOptAI.jl} bound is higher than the load shedding induced by random sampling.}\label{fig:bound}
\end{center}
\end{figure}

Sample test results are illustrated in Fig~\ref{fig:bound}, which shows the \btt{MathOptAI.jl} lower bound, vs the load shedding that resulted from random sampling of the loads and NN input parameters. Clearly, \btt{MathOptAI.jl} was able to provide a larger lower bound in every instance, generally much larger than the sampling-based approach. More exhaustive test results are provided in Tables \ref{tab:modI}, \ref{tab:modII}, and \ref{tab:modIII}. These tables compare the largest bound found by the sampling based approaches, for each grid and for each NN model, with the \btt{MathOptAI.jl} bounds. Once again, the optimized load shedding bounds are much larger than the bounds provided by a random sampling routine. We also observe a monotonic increase in load shedding as we move from \ref{tab:modI} to \ref{tab:modII} to \ref{tab:modIII} (i.e., snapping lines to their binary values tends to cause more load shed, and replacing the convex relaxation with the nonconvex power flow equations also causes a slight increase in load shed).

\begin{table}
   \caption{Model I Results (Conic Restoration with Relaxed Lines):\\\btt{MathOptAI.jl} vs the max random sample Load Shed Bound} 
   \label{tab:modI}
   \footnotesize
   \centering
   \begin{tabular}{c|cc|cc|cc}
   \toprule\toprule
   \multirow{2}{4em}{\centering\textbf{ Layer\\Size}} & \multicolumn{2}{c}{\bf{PGLib: 14} }&\multicolumn{2}{c}{\bf{PGLib: 24} Case}&\multicolumn{2}{c}{\bf{PGLib: 118}}\\ 
   &\btt{MOAI}&RND &\btt{MOAI}&RND &\btt{MOAI}&RND\\

   \midrule 
   $\;$\textit{32} & \textbf{2.28} & 0.65 & \textbf{18.05} & 1.49 & \textbf{10.45} &  4.68\\
   $\;$\textit{128} & \textbf{2.28} & 0.7 &  \textbf{18.39} & 2.37&  \textbf{15.97} &  8.86\\
   $\;$\textit{512}& \textbf{2.96} & 1.32 & \textbf{20.23} & 2.37 & \textbf{30.57} & 11.28\\
   $\;$\textit{2048}& \textbf{2.19} & 2.05 & \textbf{18.97} & 3.9  & \textbf{35.68} & 17.61\\
\bottomrule
   \end{tabular}
\end{table}

\begin{table}
   \caption{Model II Results (Conic Restoration with Snapped Lines):\\\btt{MathOptAI.jl} vs the max random sample Load Shed Bound} 
   \label{tab:modII}
   \footnotesize
   \centering
   \begin{tabular}{c|cc|cc|cc}
   \toprule\toprule
   \multirow{2}{4em}{\centering\textbf{ Layer\\Size}} & \multicolumn{2}{c}{\bf{PGLib: 14} }&\multicolumn{2}{c}{\bf{PGLib: 24} Case}&\multicolumn{2}{c}{\bf{PGLib: 118}}\\ 
   &\btt{MOAI}&RND &\btt{MOAI}&RND &\btt{MOAI}&RND\\

   \midrule 
   $\;$\textit{32} & \textbf{2.89} & 1.99 & \textbf{20.79} & 5.76 & \textbf{28.18} & 23.07\\
   $\;$\textit{128} & \textbf{2.59}&  1.72 & \textbf{20.44} & 6.37&  \textbf{35.77}&  28.88\\
   $\;$\textit{512}& \textbf{2.97} & 1.9 &  \textbf{20.79} & 6.79 & \textbf{38.74}  &27.21\\
   $\;$\textit{2048}& \textbf{2.28} & 2.05 & \textbf{18.98} & 7.3 &  \textbf{38.68}&  27.22\\
\bottomrule
   \end{tabular}
\end{table}

\begin{table}
   \caption{Model III Results (AC Restoration with Snapped Lines):\\\btt{MathOptAI.jl} vs the max random sample Load Shed Bound} 
   \label{tab:modIII}
   \footnotesize
   \centering
   \begin{tabular}{c|cc|cc|cc}
   \toprule\toprule
   \multirow{2}{4em}{\centering\textbf{ Layer\\Size}} & \multicolumn{2}{c}{\bf{PGLib: 14} }&\multicolumn{2}{c}{\bf{PGLib: 24} Case}&\multicolumn{2}{c}{\bf{PGLib: 118}}\\ 
   &\btt{MOAI}&RND &\btt{MOAI}&RND &\btt{MOAI}&RND\\

   \midrule 
   $\;$\textit{32} & \textbf{2.89} & 1.99 & \textbf{20.79} & 5.78 & \textbf{28.18} & 23.07\\
   $\;$\textit{128} & \textbf{2.59} & 1.72 & \textbf{20.44} & 6.37 & \textbf{35.77} & 28.88\\
   $\;$\textit{512}& \textbf{2.97} & 1.9 &  \textbf{20.79} & 6.8  & \textbf{39.13} & 27.21\\
   $\;$\textit{2048}& \textbf{2.35} & 2.05 & \textbf{18.97} & 7.3 &  \textbf{38.68} &  27.22\\
\bottomrule
   \end{tabular}
\end{table}

\subsection{Numerical Solution Times}

Despite its size and complexity, \eqref{eq: max_daul} was solved in times ranging from $\sim$1 second (for the smallest NNs) to $\sim$30 seconds (for the largest NN on the 118 bus systems). More exhaustive timing results are provided in Table \ref{tab:timing}. In this table, for each power grid and each NN, we present the \btt{MathOptAI.jl} solve time, and then the combined solve time of \btt{MathOptAI.jl} plus the AC feasibility restoration (i.e., Model III). We show this combination of solve times since in actual verification deployment, the conic restorations solves (Models I and II) are not needed; they simply provide useful benchmark bounds. As demonstrated by Table \ref{tab:timing}, the AC feasibility restoration models tend to solve, on the median, 100x faster than the \btt{MathOptAI.jl}, since they do not include a NN in the optimization problem. Finally, we note that these solve time numbers were somewhat sensitive to GPU responsiveness during \btt{MathOptAI.jl} solves.

\begin{table}
   \caption{Solve Times (s):  \btt{MathOptAI.jl} and AC Restoration (ACR)} 
   \label{tab:timing}
   \footnotesize
   \centering
   \begin{tabular}{c|cc|cc|cc}
   \toprule\toprule
   \multirow{2}{4em}{\centering\textbf{ Layer\\Size}} & \multicolumn{2}{c}{\bf{PGLib: 14} }&\multicolumn{2}{c}{\bf{PGLib: 24} Case}&\multicolumn{2}{c}{\bf{PGLib: 118}}\\ 
   &\btt{MOAI}&\btt{M}+ACR &\btt{MOAI}&\btt{M}+ACR &\btt{MOAI}&\btt{M}+ACR\\

   \midrule 
   $\;$\textit{32}  & 1.7  & 1.71 & 3.02 & 3.06 &  7.6  &  8.03\\
   $\;$\textit{128} & 1.21 & 1.22 & 0.98 & 1.0  & 11.06 & 15.25\\
   $\;$\textit{512} & 1.85 & 1.88 & 9.41 & 9.43 & 11.2  & 11.42\\
   $\;$\textit{2048}& 8.7  & 8.76 & 7.08 & 7.12 & 30.83 & 31.04\\
\bottomrule
   \end{tabular}
\end{table}

\subsection{Load Shedding Patterns}

\begin{figure}
\begin{center}
\includegraphics[width=0.9\columnwidth]{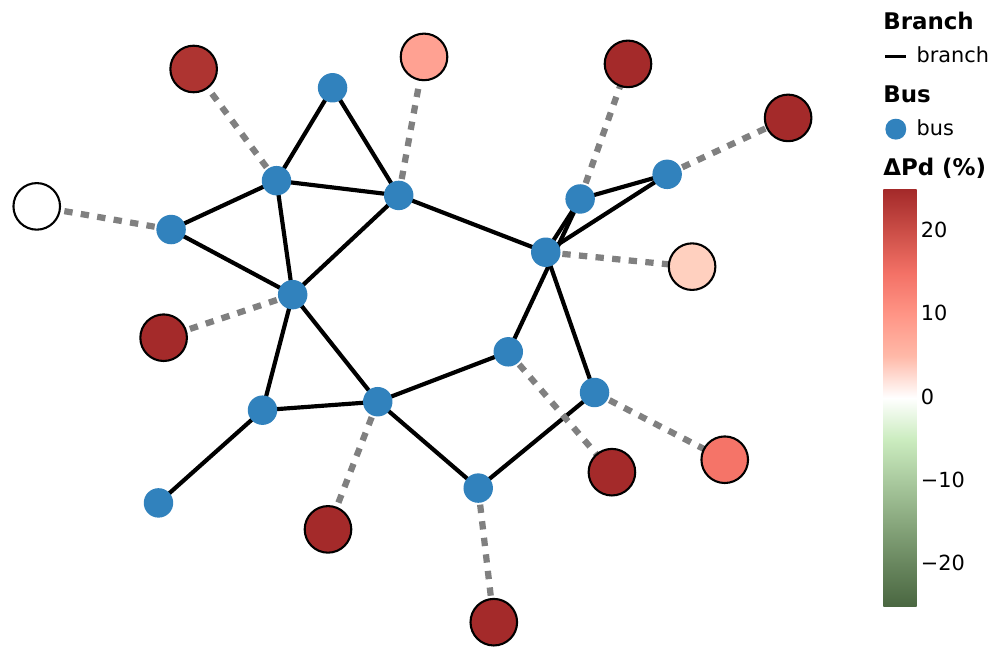}
\caption{Shown is the change in \textit{active} power load, relative to the base load, when \btt{MathOptAI.jl} solves \eqref{eq: max_daul}. As depicted, most of the active power loads increase to their upper limits, given by \eqref{eq: gamma_bounds}.}\label{fig:powerplot_pd}
\end{center}
\end{figure}

\begin{figure}
\begin{center}
\includegraphics[width=0.9\columnwidth]{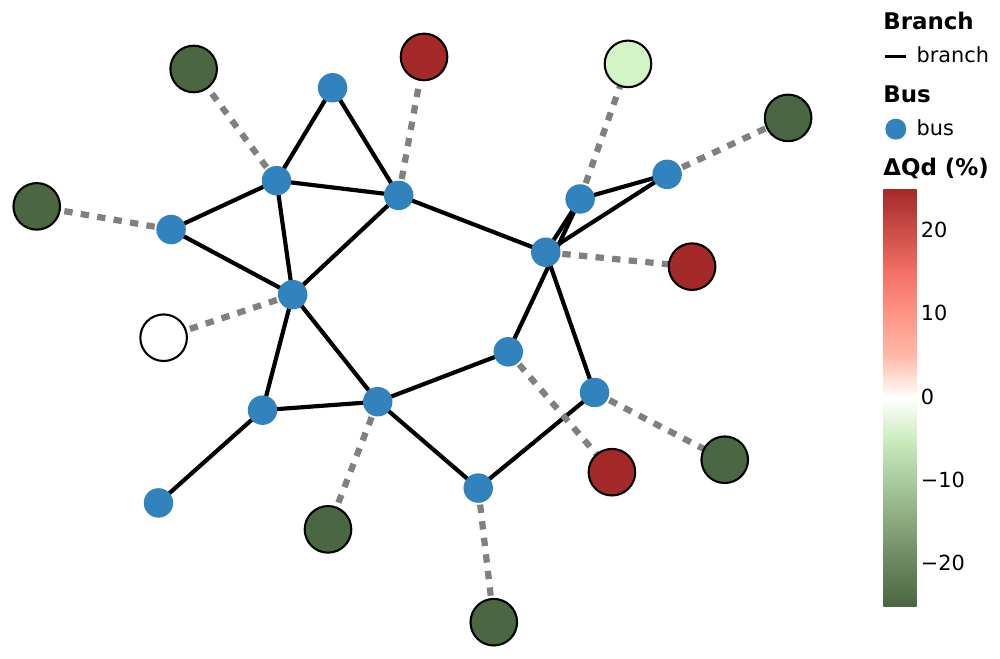}
\caption{Shown is the change in \textit{reactive} power load, relative to the base load, when \btt{MathOptAI.jl} solves \eqref{eq: max_daul}. As depicted, most of the reactive power loads decrease to their lower limits, given by \eqref{eq: gamma_bounds}.}\label{fig:powerplot_qd}
\end{center}
\end{figure}

In this section, we demonstrate some of the observed loading patterns which emerge when \btt{MathOptAI.jl} optimized for the maximal load shedding. In Figs. \ref{fig:powerplot_pd} and \ref{fig:powerplot_qd}, which were generated using \btt{PowerPlots.jl}~\cite{rhodes2025powerplots}, we show the optimal network loading identified by \btt{MathOptAI.jl} for the 14-bus grid and the largest NN (2048 activation functions per layer). To recall, in the optimization formulation, load is allowed to vary by $\pm 25\%$. Fig.~\ref{fig:powerplot_pd} shows the optimal active power loading change (relative to the base loading), while Fig.~\ref{fig:powerplot_qd} shows the optimal reactive power loading change. Notably, the optimizer tended to \textit{maximize} active power loading and \textit{minimize} reactive power loading. This result is counterintuitive, since increased $Q$ loading tends to correlate with load shedding. However, we offer the following nuanced explanation:
\begin{itemize}
\item \textbf{Increased ${P}$:} Active power loading was probably increased by the optimizer to increase overall load shedding (in accordance with the objective \eqref{eq:load_objective}): more active loading $\rightarrow$ more load to shed.
\item \textbf{Decreased ${Q}$:} More load will also be shed when more lines are turned off (i.e., turned probabilistically down). As reactive power is turned down, the OPS model (which the NN is trained to emulate) can turn off more lines (which it wants to do, to minimize risk à la \eqref{eq:obj}) while still serving an equivalent amount of active power (which is its goal). Less $Q$ $\rightarrow$ turn off more lines $\rightarrow$ serve same P. Thus, $Q$ loading and line status correlate positively. The optimizer probably turned down $Q$ to turn down line statuses, thus forcing the redispatch algorithm to shed more load overall.

\end{itemize}



\section{Conclusions}\label{conclusion}

In this paper, we trained a NN model to make optimal line switching decisions in an AC power grid. Next, we considered the following question: in actual deployment, what is the worst-case load shedding that might result from switching decisions suggested by this NN? We posed this as a bilevel optimization problem, where the ``attacker" was modeled by a NN making switching decisions, and the ``defender" was modeled by an AC-Redispatch function which minimizes load shedding. To solve this problem, we convexified and dualized the inner dispatch problem, combining it with the attacker maximization problem, resulting in a tractable optimization problem. We solved this problem with \btt{MathOptAI.jl}. The resulting dual formulation was numerically challenging to solve, but we could still solve over the largest NN models (with $\sim$10M model parameters) in $\sim$30s. 

Overall, the proposed framework shows promising performance for tightly lower bounding the worst-case load shedding. We believe this framework could be applied to much larger power grid tests cases, assuming NNs could be trained to solve switching decisions in these networks. However, the numerical robustness of the model would need to be improved to ensure reliable convergence. Future work will target better convergence, and it will also target combining this work with NN convex relaxations in order to upper bound the worst-case load shedding problem.

\appendices

{\section{Power Flow Formulations}\label{AppA}}
Using lifted SOC voltage variables, we present the power flow equations. Parameters include the line conductance ${g}_{ij}$ line susceptance ${b}_{ij}$, shunt conductance ${g}_i$, shunt susceptance ${b}_i$, transformer ratio magnitude $|{t}_{ij}|$, real part of the transformer ratio ${t}^R_{ij}$, and imaginary part of the transformer ratio ${t}^I_{ij}$.  The resulting power flow equations, which are defined for each direction, are linear in the voltage variables:
\begin{subequations}
\begin{align}
    \begin{split}
        P_{ij}^L &=   
            \frac{{g}_{ij} + {g}_{i}}{|{t}_{ij}|^2} W^{Fr}_{ij} 
            + \! \frac{-{g}_{ij} {t}^R_{ij} \!+\! {b}_{ij} {t}^I_{ij}}{|{t}_{ij}|^2} W^R_{ij} \\ 
            & 
            + \! \frac{-{b}_{ij} {t}^R_{ij} \!-\! {g}_{ij} {t}^I_{ij}}{|{t}_{ij}|^2} W^I_{ij}
        \quad \forall ij \in \mathcal{L} \label{eq:soc_p_power_flow_fr}
        \end{split}\\
    \begin{split}
        P_{ji}^L &=    
            \left( {g}_{ij} + {g}_{j} \right) W^{To}_{ij} 
            + \! \frac{-{g}_{ij} {t}^R_{ij} \!-\! {b}_{ij} {t}^I_{ij}}{|{t}_{ij}|^2} W^R_{ij} \\
            & 
            + \! \frac{-{b}_{ij} {t}^R_{ij} \!+\! {g}_{ij} {t}^I_{ij}}{|{t}_{ij}|^2} W^I_{ij}
        \quad \forall ij \in \mathcal{L}  \label{eq:soc_p_power_flow_to}
        \end{split}\\
    \begin{split}
        Q_{ij}^L &=    
             -\frac{{b}_{ij} + {b}_{i}}{|{t}_{ij}|^2} W^{Fr}_{ij} 
            - \! \frac{-{b}_{ij} {t}^R_{ij} \!-\! {g}_{ij} {t}^I_{ij}}{|{t}_{ij}|^2} W^R_{ij} \\ 
            & 
            + \! \frac{-{g}_{ij} {t}^R_{ij} \!+\! {b}_{ij} {t}^I_{ij}}{|{t}_{ij}|^2} W^I_{ij}
         \quad \forall ij \in \mathcal{L} \label{eq:soc_q_power_flow_fr} 
    \end{split}\\
    \begin{split}
        Q_{ji}^L &=    
             - \left( {b}_{ij} + {b}_{j} \right) W^{To}_{ij} 
            - \! \frac{-{b}_{ij} {t}^R_{ij} \!+\! {g}_{ij} {t}^I_{ij}}{|{t}_{ij}|^2} W^R_{ij} \\ 
            & 
            + \! \frac{-{g}_{ij} {t}^R_{ij} \!-\! {b}_{ij} {t}^I_{ij}}{|{t}_{ij}|^2} W^I_{ij}
         \quad \forall ij \in \mathcal{L}. \label{eq:soc_q_power_flow_to}
    \end{split}
\end{align}
\label{eq:ops_soc_powerflow}
\end{subequations}

The nonconvex AC power flow equations are given, where the binary variable $z_{ij}$ controls whether a line is on or off:
\begin{subequations}
\begin{align}
    P_{ij}^L &= z_{ij}  \Bigg( \Bigg. 
        \frac{{g}_{ij} + {g}_{i}}{|{t}_{ij}|^2} V_i^2 
        \!\! + \! \frac{-{g}_{ij} {t}^R_{ij} \!+\! {b}_{ij} {t}^I_{ij}}{|{t}_{ij}|^2} V_{i} V_{j} \cos\left(\theta_{i} \!-\! \theta_{j}\right)\nonumber \\ 
        & \!\!\!\! + \! \frac{-{b}_{ij} {t}^R_{ij} \!-\! {g}_{ij} {t}^I_{ij}}{|{t}_{ij}|^2} V_{i}  V_{j} \sin\left(\theta_{i} \!-\! \theta_{j} \right) 
    \!\! \Bigg. \Bigg) \: \forall ij \in \mathcal{L} \label{eq:ac_p_power_flow_fr}\\
    P_{ji}^L &= z_{ij}  \Bigg( \Bigg. 
        \left( {g}_{ij} + {g}_{j}\right) V_j^2 
        \!\! + \! \frac{-{g}_{ij} {t}^R_{ij} \!-\! {b}_{ij} {t}^I_{ij}}{|{t}_{ij}|^2} V_{j} V_{i} \cos\left(\theta_{j} \!-\! \theta_{i}\right)\nonumber \\
        & \!\!\!\! + \! \frac{-{b}_{ij} {t}^R_{ij} \!+\! {g}_{ij} {t}^I_{ij}}{|{t}_{ij}|^2} V_{j}  V_{i} \sin\left(\theta_{j} \!-\! \theta_{i} \right) 
    \!\! \Bigg. \Bigg) \: \forall ij \in \mathcal{L}  \label{eq:ac_p_power_flow_to}\\
    Q_{ij}^L &= z_{ij}  \Bigg( \Bigg.
         -\frac{{b}_{ij} + {b}_{i}}{|{t}_{ij}|^2} V_i^2 
         \!\! - \! \frac{-{b}_{ij} {t}^R_{ij} \!-\! {g}_{ij} {t}^I_{ij}}{|{t}_{ij}|^2} V_{i} V_{j} \cos\left(\theta_{i} \!-\! \theta_{j}\right)\nonumber \\ 
        & \!\!\!\! + \! \frac{-{g}_{ij} {t}^R_{ij} \!+\! {b}_{ij} {t}^I_{ij}}{|{t}_{ij}|^2} V_{i} V_{j} \sin\left(\theta_{i} \!-\! \theta_{j}\right) 
    \!\! \Bigg. \Bigg) \: \forall ij \in \mathcal{L} \label{eq:ac_q_power_flow_fr} \\
    Q_{ji}^L &= z_{ij}  \Bigg( \Bigg.
         - \left( {b}_{ij} + {b}_{j} \right) V_j^2 
         \!\! - \! \frac{-{b}_{ij} {t}^R_{ij} \!+\! {g}_{ij} {t}^I_{ij}}{|{t}_{ij}|^2} V_{j} V_{i} \cos\left(\theta_{j}\!-\!\theta_{i}\right)\nonumber \\ 
        & \!\!\!\! + \! \frac{-{g}_{ij} {t}^R_{ij} \!-\! {b}_{ij} {t}^I_{ij}}{|{t}_{ij}|^2} V_{j} V_{i} \sin\left(\theta_{j}\!-\!\theta_{i}\right) 
    \!\! \Bigg. \Bigg) \: \forall ij \in \mathcal{L}. \label{eq:ac_q_power_flow_to}
\end{align}
\label{eq:ops_ac_powerflow}
\end{subequations}

{\section{Implementation Details}\label{AppB}}
There were two primary computational challenges in collecting the test results. First, the dual maximal load shedding formulation \eqref{eq: max_daul} tended to be hard for the IPOPT optimizer to solve. To overcome this, $(i)$ we set \btt{mu\_init} to $10^{-8}$ in all IPOPT solves, $(ii)$ we used a convergence tolerance of $5^{-3}$ (the solution was subsequently ``cleaned up" in the ``conic restoration" step), $(iii)$, we dropped the angle voltage limit constraint (as it didn't seem to affect the solutions), and $(iv)$ we excluded the binary variable $z_{ij}$ in SOC constraint \eqref{eq:ops_complex_thermal_limit} (this also tended to not affect the solution, since the binary was also driving line voltage differentials to 0). Finally, in the reformulation of the AC-Redispatch problem, shunt power injections were modeled as continuous relaxations (e.g., $0 \le P_i^S\le g_iW_{ii}$) rather than as McCormick relaxations \eqref{eq:ops_soc_shunt_relaxation}.

The second challenge was related to rejection sampling on the 118-bus network. When we sampled loads, fed the resulting input vector into the largest NNs, computed line status probabilities, and snapped the binaries, many lines could be shut off, leading to infeasible restoration problems. This numerical infeasibility was an artifact of the model representation, so we rejected all samples that couldn't be solved. Future work would re-formulate the model to avoid such numerical infeasibility.


\bibliographystyle{IEEEtran}
\bibliography{IEEEabrv,ref}

\end{document}